\documentclass[prl,aps,showpacs,floatfix,byrevtex,amsmath,superscriptaddress,amssymb,twocolumn]{revtex4}

\usepackage{graphics}
\usepackage{amssymb}

\begin{document}


\title{New criteria for the equation of state development: \\
Simple model fluids}

\author{Yurko Duda and Pedro Orea}

\affiliation{\small Programa de Ingenier\'{\i}a Molecular, Instituto Mexicano del Petr\'{o}leo, Eje Central 152,
07730 M\'{e}xico D.F., M\'exico; e-mail: yduda@lycos.com}


\begin{abstract}

\noindent

Recently we have proposed (J. Chem. Phys. 128 (2008) 134508) a new rescaling of fluid density $\rho $ by its
critical value $\rho_c^{2/3}$ to apply the corresponding states law for the attractive Yukawa fluids study.
Analysis of precise simulation results allows us to generalize this concept to the case of simple fluids with
different interparticle interactions, like Mie (n,m) and Sutherland pair potentials. It is shown, that there is
a linear relationship between the critical pressure and critical temperature, as well as the critical density
and inverse critical temperature for these frequently used pair potentials. As a consequence, the critical
compressibility factor of these model fluids is close to its universal value measured experimentally for
different real substances.

\end{abstract}

\maketitle


Equation of state (EOS) is of crucial importance to predict or correlate fluid thermodynamic properties
\cite{valder1,fazl1,kudel1}. Since van der Waals proposed the first version of his EOS \cite{vdw}, many
modifications have been proposed, and today, there are numerous more accurate EOSs reported in the literature
for describing the behavior of simple fluids and fluid mixtures. An improvement in the accuracy of analytical
EOSs for the modeling of experimental data of real fluids is usually achieved, not by improving theoretical
base, but by introducing empirical temperature functionalities in their parameters. Obviously, such empirical or
semi-empirical approaches, beside to require a lot of experimental \cite{valder1,fazl1} or simulation data
\cite{niko1,nasrad}, forbidden molecular interpretation of results. On the other hand, theoretically based EOSs
are very general and applicable for the entire family of systems obeying the given interparticle potential model
\cite{nezb0,paric1,szhou3,kal1,tang0,beta1,zhou1,kolaf1,haro1,shok1}.

 The most apparent progress toward EOS was made by applying principles of statistical mechanics.
 Integral equation \cite{kal1,tang0}, thermodynamic perturbation \cite{beta1,zhou1,kolaf1,haro1,shok1,szhou3},
 and mean field theories \cite{volker} are the most common theoretical approaches to
 develop  analytical EOS for simple fluids. The advantage of theoretically based EOS over their empirical
 counterparts is that they can be further improved by testing against computer simulations and later be
 used as a reference for perturbation theory calculations of bulk and interfacial properties of
 more complex fluids such as polymers and associating systems (i.e., hydrogen-bonding fluids)
 \cite{alej1,jack1,fu1,rzys1,cum1,tuin1}.

  Although last decades significant progress has been achieved in the development of theoretically based
  expressions for EOS, there is still a lack of equations that described PVT properties of simple fluids in wide
  range of thermodynamic parameters. Such precise analytical expression may be obtained if one has a set of
  universal criteria (as for example, corresponding states law \cite{gugg1,kisel1,orea1}) or/and apply self-consistent
  integral equations \cite{scoza}.

Recently, we have applied the corresponding states law for the hard core attractive Yukawa (HAY) fluid
\cite{orea1}, and revealed that after a certain rescaling of fluid number density $\rho $ the reduced surface
tension and pressure map onto the master curves. To make such analysis we performed canonical Monte Carlo (MC)
simulations with relatively high precision \cite{duda7}. The goal of the present work is to show that the master
curve for the supercritical reduced pressure is not restricted to the HAY fluid and is universal for other
simple nonelectrostatic fluid pair potentials. Namely, in this communication we consider the following widely
used pair potentials \cite{duda7,okumura,camp,noro,pana1,john2}:

\vspace{1cm}

 1)  Mie (n,m) potential,

\begin{equation}
\label{Mie}
 u (r)=  \epsilon \Big( \frac{n}{n-m}\Big ) \Big(\frac{n}{m}\Big )^{m/(n-m)} \Big [\Big (\frac{1}{r}\Big )^n -
 \Big (\frac{1}{r} \Big )^m
 \Big ]
\end{equation}
\vspace{1cm}

2) Sutherland potential,

\begin{equation}
\label{suth}
u(r)=\left\{\begin{array}{ll}\infty, & \mbox{ if $r < 1,$}\\
    -\epsilon \Big (\frac{1}{r} \Big )^\gamma, & \mbox{ if $ r\geq 1,$}
     \end{array} \right.
\end{equation}

and,
\vspace{1cm}

3) HAY potential,

\begin{equation}
\label{yuka}
u(r)=\left\{\begin{array}{ll}\infty, & \mbox{ if $r<1,$}\\
-\epsilon \exp[-\kappa (r-1)]/r, & \mbox{ if $ r\geq
    1,$}
 \end{array} \right.
\end{equation}

where $\epsilon$ is the potential well depth, $\kappa$ defines the range of the potential, and $r$ is an
interparticle distance reduced by particle diameter, which is chosen to be a unit of length.

Note, that last two potentials, eqs. (\ref{suth}) and (\ref{yuka}), have hard sphere repulsive part, while the
family of Mie potentials (which includes widespread used Lennard-Jones (LJ) pair potential, when $n=12$, and
$m=6$) reproduces a soft repulsive part at short distances and a smooth dispersion part at intermediate and long
distances. In other words,  the distinct nature of the potentials described above is obvious.

 In this work we mainly use the simulation data of critical parameter values reported in the literature
 \cite{duda7,okumura,camp,noro,pana1}. Some new critical data shown in Table 1 were estimated
by fitting the MC simulation data for coexisting liquid $\rho_L$ and vapor $\rho_V$ densities \cite{reyes} to
the rectilinear diameter law \cite{john2,miguel}

\begin{equation}
\label{first} \frac{\rho_L + \rho_V}{2} = \rho_c + A |T/T_C -1 |,
\end{equation}

 and to the following equation

\begin{equation}
\label{second} \rho_L - \rho_V = B |T/T_C -1 |^{\beta},
\end{equation}

where $A$ and $B$ are the coefficients of the leading terms in the general Wegner expansions, and $\beta =
0.325$ is the universal value of critical exponent in the case of short-ranged potentials \cite{john2,miguel},
that belong to the Ising universality class \cite{BETA,wilding}. Up to four pairs of densities, $\rho_L (T_i)$
and $\rho_V(T_i)$, at temperatures $T_i$ close to its critical value, have been used in Eqs. (\ref{first}) and
(\ref{second}) to estimate the critical parameters $\rho_c$ and $T_c$; averaging of the six values (combination
of the four pairs of density) has provided us the mean values and uncertainties of the critical values presented
in Table 1. It is well accepted \cite{BETA} that such kind of estimation gives the same values of critical data
as obtained by other more sophisticated simulation techniques \cite{john2,wilding}.

The critical pressures were estimated on the base of the Clasius-Clapeyron equation. Besides, we have performed
canonical MC simulations to calculate supercritical pressures for the fluids with pair potentials defined above.
All details of the simulation can be found elsewhere \cite{duda7,orea1}; as usual, additional simulation runs of
different size systems and/or with different length of $r_{cut}$ have been performed to verify that our results
do not depend on these parameters.

   First of all, let us consider the relation between the critical parameters of the liquid-vapor phase
   transition
   of the fluids with different pair potentials. As seen in Fig.1, the critical pressure $P_c$ is a
linear function of the critical temperature $T_c$ for the different potentials considered. Note that family of
Mie potentials is represented by more than $20$ points which correspond to the different combinations of
parameters $n$ and $m$ in eq.(\ref{Mie}). Besides, in Fig.1 there are some points of the slightly modified Mie
potentials, which were studied in Ref. \cite{pana1}; namely, the Buckingham exponential-6 $(Bexp6)$, and the
LJ($\alpha - 6$) potentials.  It is important to remark, that the linear function $P_c=P_c(T_c)$ is nearly the
same as was reported in Ref. \cite{orea1} for HAY fluid potential with different values of $\kappa$.

As shown in Fig.2, the critical densities $\rho_c$ are linearly dependent on the inverse critical temperature.
Some of the data on the $\rho_c = \rho_c (1/T_c)$ curve (for example, the data of Ref. \cite{pana1}, which were
not used for the linear fitting) are located slightly out of the adjusting solid curve, although presenting
linear dependence too. It is important to note, that the two linear functions shown in Figs. 1 and 2 allow to
obtain the value of the critical compressibility factor, $Z_c = P_c/(\rho_c T_c)$, around $0.30 \pm 0.01$ for
all the model fluid considered here. This value of $Z_c$ is close to the universal value reported for different
real substances \cite{gugg1}, and usually is being used for the critical pressure estimation. Prediction of
$Z_c$ obtained from various theoretical approaches \cite{kal1,zhou1,haro1,shok1,volker,jack1,alej1} are far from
this value because, as well known, none of them are sufficiently precise in the critical region.

  Finally, in Fig. 3 the reduced pressure, $P_R = P/P_c$, as a function of the rescaled density,
  $\rho_r = \rho / \rho_c^{2/3}$, is presented for the four different potentials.
  Such rescaling of density has been proposed and tentatively justified in our previos work
  \cite{orea1}, where the extended corresponding states law has been applied for the description of the
  HAY fluid properties. As can be seen in the figure, for the three reduced temperatures
  considered, all four model fluids match on the same master curves at each value of $T_R = T/T_c$.
  The best agreement of the simulation data is reached at the lowest temperatures and densities;
  only slight discrepancy (almost within the error bars) is observed for the highest densities (or pressures).


In summary, novel criteria for estimating the critical parameters and reduced equation of state of simple fluid
model potentials is proposed. Namely, the rule of linear dependence of critical pressure on critical
temperature, as well as critical density on inverse critical temperature is tested for a number of potentials.
Such rule gives critical compressibility factor around $0.30 \pm 0.01$ which agrees well with experimental data
of some real gases. Besides, new re-scaling of simple fluid density, $\rho_r = \rho / \rho_c^{2/3}$ allows one
to describe the reduced pressure by a single reduced isotherm. We believe, that the proposed universality
relations will help to improve old theoretical approximation in the field of simple fluid modeling, and develop
new ones. Moreover, it would be interesting to analyze the available experimental PVT data with the idea to
verify the applicability of the corresponding states rules presented here. More extensive computer simulations
are necessary to investigate their applications for the interfacial properties of simple model fluids. This
issue will be considered in future reports from our laboratory \cite{reyes}.

\section{Acknowledgments}

We gratefully acknowledge the financial support of the Instituto Mexicano del Petr\'oleo under the Projects
$D.31519/D.00406$.

\newpage

.

\section*{Figure Captions}



\noindent \hspace {1cm} { {\bf Fig. 1} {\small {Relation between critical pressure $P_c$ and temperature $T_c$
for different model potentials: HAY with $1.5<\kappa<7$ \cite{orea1}, LJ($\alpha - 6$) with $10<\alpha<22$
\cite{pana1}, $Mie(n-6)$ with $7<n<32$ \cite{okumura}, $Mie(2n-n)$ with
$6<n<18$ \cite{noro}, and $Bexp6$ with $12<\alpha<22$ \cite{pana1}.              }\\

\noindent \hspace {1cm} { {\bf Fig. 2} {\small {Relation between critical number density $\rho_c$ and inverse
critical temperature $T_c^{-1}$ of the model potentials considered in fig.1, and Sutherland potential with
$\gamma = 4$, and $6$
\cite{camp} }}\\

\noindent \hspace {1cm} { {\bf Fig. 3} {\small {Reduced pressure, $P_R = P/P_c$ as a function of reduced fluid
density
$\rho_r = \rho / \rho_c^{2/3}$. }}\\

\vspace{0.5cm}

\centering \resizebox{0.5\textwidth}{0.35\textheight}{\includegraphics{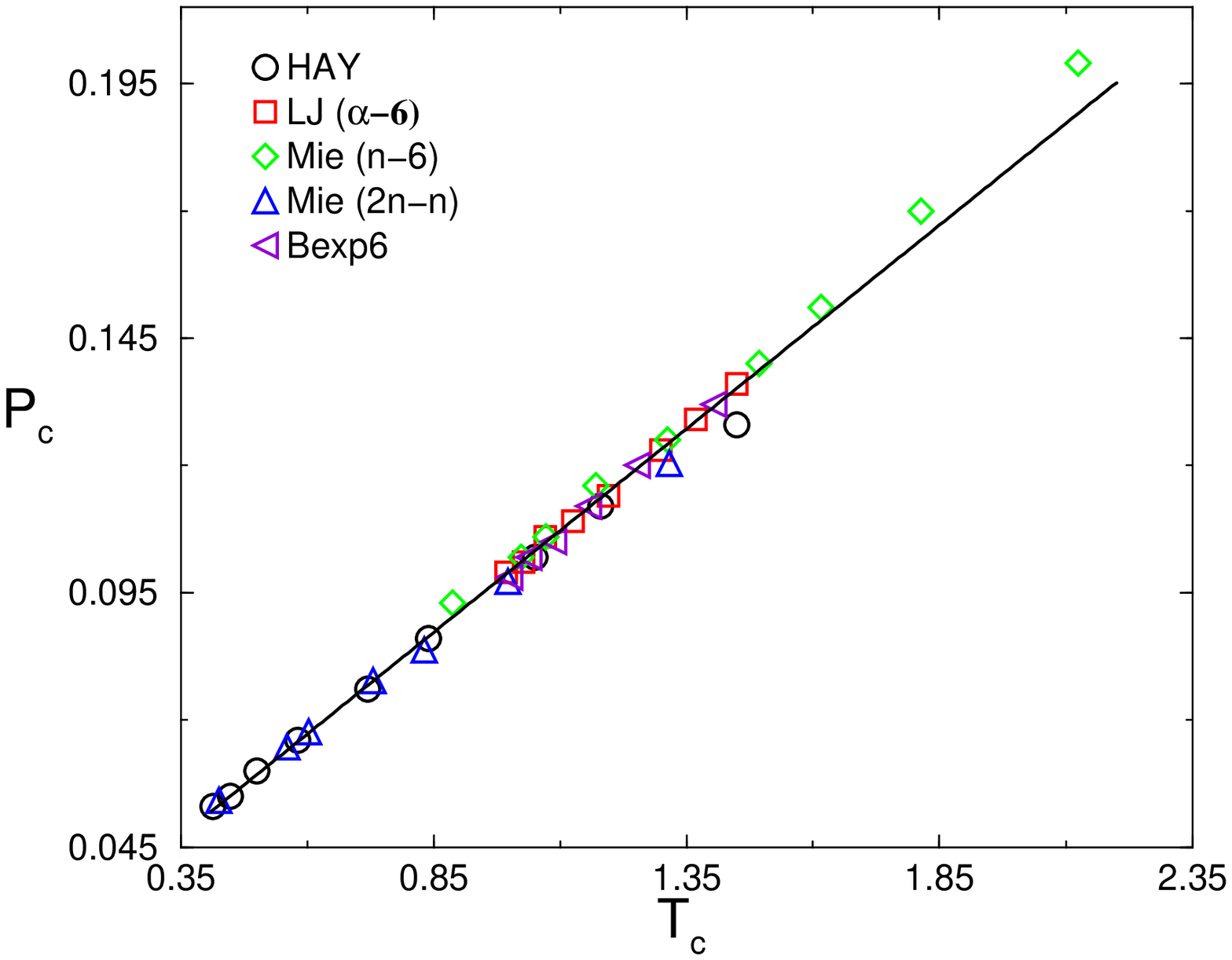}} \vspace{1.0cm}
\centerline{Fig. 1}
\newpage

\resizebox{0.5\textwidth}{0.35\textheight}{\includegraphics{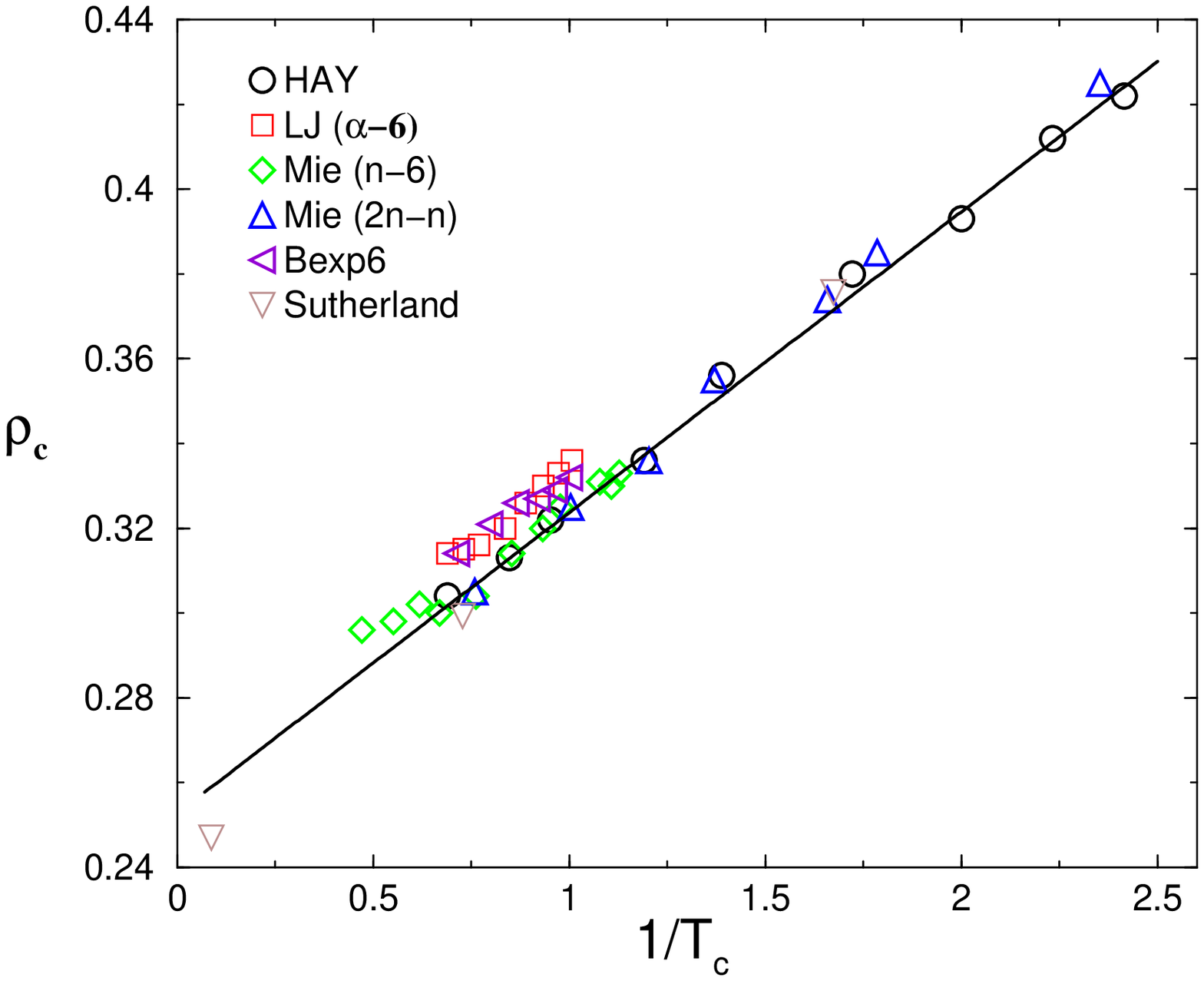}} \vspace{1.0cm} \centerline{Fig. 2}

\resizebox{0.5\textwidth}{0.35\textheight}{\includegraphics{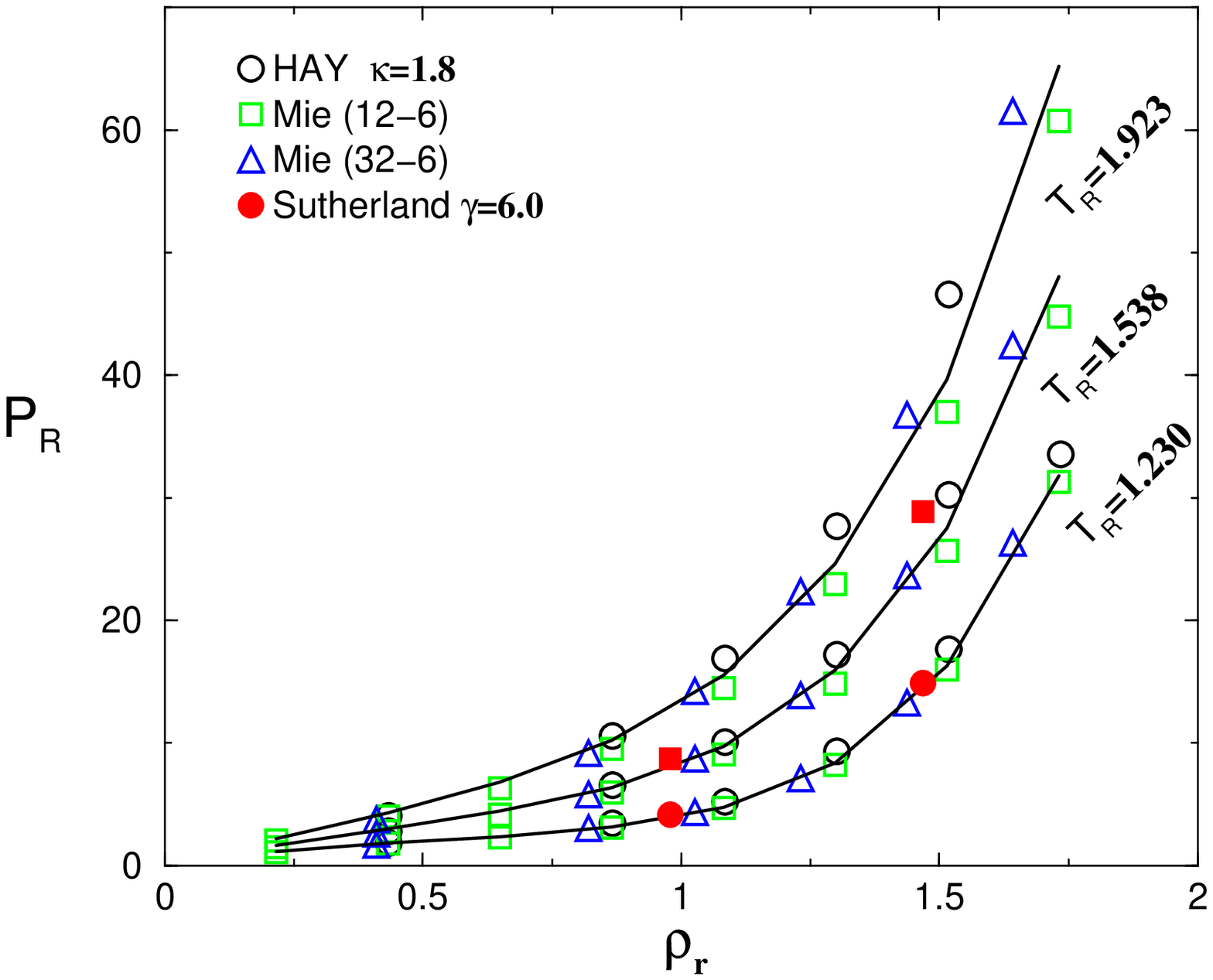}} \vspace{1.0cm} \centerline{Fig. 3}

\begin{table}
\caption{Critical parameters of the four model potentials obtained from the MC simulations, and used for the
$P_R = P_R(\rho_r)$ function calculations (see Fig.3). The last column presents the cutoff distance $r_{cut}$
applied for each pair potential.}

\label{table1}

\begin{tabular}{cccccccc}
\hline\hline \hspace{1.5cm}$Potentials$\hspace{0.8cm} & $T_c$\hspace{0.8cm}  & $\rho_c$\hspace{0.8cm}
&  $P_c$\hspace{1.cm}  &  $Z_c$ &  $r_{cut}$ \hspace{0.7cm}  \\

\hline

 HAY $\kappa = 1.8$  (Ref. \cite{duda7})  & $1.180_5$  & $0.313_6$  &  $0.112_2$   &  $0.303_7$ & 5.0  \\
 Mie (12-6) (LJ fluid) \cite{reyes} & $1.290_9$   & $0.314_6$  &  $0.118_3$   &  $0.291_8$ &  5.5  \\
 Mie (32-6)   \cite{reyes}          & $0.865_4$   & $0.340_6$  &  $0.090_3$   &  $0.306_7$ & 4.0   \\
 Sutherland  ($\gamma =6.0$) \cite{reyes} & $0.595_5$  & $0.365_5$  &  $0.066_2$   &  $0.303_6$ & 4.0   \\

\end{tabular}
\end{table}

\end{document}